\documentclass[conference]{IEEEtran}
\usepackage{graphicx}
\usepackage{float}
\usepackage{textcomp}
\usepackage{setspace}
\usepackage{array}
\usepackage{amsmath}
\usepackage{subfigure}
\begin{document}
\title{ Two novel metrics for determining the tuning parameters of the Kalman Filter}

\author{Manika Saha \hspace{1cm} Bhaswati Goswami  \hspace{1cm} and \hspace{1cm} {Ratna Ghosh \thanks{Address for Correspondence:-
Dr. Ratna Ghosh,
Department of Instrumentation and Electronics Engineering,
Jadavpur University 2nd Campus, Salt Lake
Kolkata - 700 098, India.
E-mail: {ratna\_g}@hotmail.com}} \\
Department of Instrumentation and Electronics Engineering \\ Jadavpur University 2nd Campus\\ Block LB, Salt Lake \\
Kolkata  700 098 \\ INDIA}

\date{}
\maketitle
\thispagestyle{empty}

\begin{abstract}
The Kalman filter (KF) and the extended Kalman filter (EKF) are well established techniques for state estimation.
However, the choice of the filter tuning parameters still poses a major challenge for the engineers \cite{Ananthasayanam2011}.
In the present work, two new metrics have been proposed for determining the filter tuning parameters on the basis of the innovation covariance.
This provides a metric based offline method usable for predicting the actual filter RMSE performances for a particular application and thus, for the selection of suitable combination(s) of the filter tuning parameters in order to ensure the design of a KF or an EKF having a balanced RMSE performance.
\end{abstract}


\section{Introduction}
The Kalman filter (KF) and its extension for non-linear systems using the linearized system matrices, the extended Kalman filter (EKF),
are well established techniques for state estimation which has important applications in various fields like control, monitoring and/or
 fault detection of various systems and processes.  Both the KF and the EKF provide simple yet effective methods for state estimation
 by accounting for the unmodeled dynamics and measurement noises using the system covariance matrices. The Kalman estimation problem
 essentially involves the computation of the Kalman filter gain using the estimated system uncertainty and noise covariance matrices.
 However, it is well-known that a major drawback of these filters is that they require an apriori knowledge about the process and the
 measurement noise statistics. In most practical applications, the exact information about these statistics is unavailable and so,
 these covariance matrices, referred to as the filter tuning parameters, have to be supplied by the designer using some {ad-hoc}
  procedures \cite{Ananthasayanam2011,Cordova2009}. Thus, the choice of the filter tuning parameters still poses a major challenge for the
  engineers \cite{Ananthasayanam2011,Ananthasayanam2002}.

 Several researchers have tried various methods and approaches for choosing the filter tuning parameters to rationalize the {ad-hoc} nature of the choice.
 Conventional methods for tuning like downhill simplex numerical optimization algorithm \cite{Powell2002} as well as  modern techniques like Neural
 Network (NN), genetic algorithm and fuzzy logic based approaches \cite{Korniyenko2005, Clements2000, Rahbari2002} have been used for the KF tuning
 problem. Rosendo et al. \cite{Rosendo2007} designed a self-tuned Kalman filter and compared it with the conventional running average and a
 conventionally tuned Kalman filter. These filters were used in the low-pass filtering stage required in some active power filter algorithms and
 their results showed that though all the three methods perform well, but the self-tuned Kalman filter reacts faster under transient conditions.
 A method using the normalization of the system matrices has been used for the choice of the covariance matrices for the online determination of
 rotor position and speed of a permanent-magnet synchronous motor in \cite{Bolognani2003}. However, none of these methods have been able to provide
 a deterministic method for selecting the filter covariances.
 Some researchers have focussed on innovations in order to address the filter tuning issue.  One method involves the use of the estimated
 autocovariance of the output innovations  to compute a least-squares estimate of the noise covariance matrices \cite{Aakesson2007}.
 Kailath \cite{Kailath1968} also proposed that the innovations be measured and their mean and covariance be approximated using statistical
 methods in order to verify the KF performance and to adjust the KF parameters to improve the performance of the state estimation.
 He further stated that if the mean and covariance of the innovations are not as expected then it indicates that the choice of any
 or all of the system matrices as well as the covariances is/are incorrect. An innovation based cost function for KF, termed as the
 normalized innovation squared (NIS),  has been suggested in \cite{Shalom2001} which uses the UMPITS \cite{Paramonova2006} to find out
 whether the particular choice of the tuning parameters can assure convergence of the filter. This cost is defined in terms of the
 innovation $q_k$ and the innovation covariance $S_k$ as

\begin{equation} \label{eqn:NIS}
J=\frac{1}{N}\sum {q_k^T}(S_k)^{-1}q_k
\end{equation}

Under the hypothesis that the filter is consistent, the NIS has a
chi-square distribution with $n_z$ degrees of freedom, where $n_z$
is the dimension of the measurements. One of the limitations of
this method is that it has to be tested online and so it cannot be
used for
 predictions of suitable choices of the filter tuning parameters but can only be used for verification of dimensional consistency of the filter.

In the present paper, a metric based predictive method, for the selection of suitable combination(s) of filter tuning parameters,
has been proposed in order to ensure the design of a Kalman filter having a judiciously balanced performance in terms of robustness and
sensitivity. For this purpose, two novel performance indices (metrics) have been suggested which can be used to predict and/or compare the
quality of the practically obtained RMSE performances. In Section II, the new performance indices have been derived and their significance
has been discussed. A realistic problem, used to demonstrate the effectiveness of the proposed metrics, is stated in detail in Section III,
while the corresponding simulations and results are stated in Section IV.
Conclusions are stated in Section V.

\section{Novel Performance Indices for the choice of Filter Tuning Parameters}

A linear (nonlinear) discrete time (stochastic) system may be described using linear(ized) state and observation equations, at a particular time instant $k$, as follows:
\begin{eqnarray}
  \nonumber {x_{k+1}}&=&{F_k}{x_k}+{G_k}{u_k}+{w_k} \\
  {y_k}&=&{H_k}{x_k}+{v_k} \label{eqn:1}
\end{eqnarray}
Here, $u_k$ is the known input, while $w_k$ and $v_k$ are the state and measurement noises. Both of these are zero-mean, uncorrelated white noises with their
 covariances being $Q_k$ and $R_k$ respectively. $F_k$, $G_k$ and $H_k$ are the state transition, input and measurement matrices respectively.

The Kalman filter, which estimates the state vector $x_k$ from the measurements $y_k$ in an optimal sense, can be expressed
as a set of sequential equations for the apriori state estimate and error covariance, $\hat{x}_k^-$ and ${P}_k^-$,
the innovation and its covariance, $q_k$ and $S_k$, the Kalman Gain, $K_k$, and the aposteriori state estimate and error covariance $\hat{x}_k^+$ and ${P}_k^+$ as follows:
\begin{eqnarray}\label{eqn:cov}
\hat{x}_k^-  & = & F_{k-1} \hat{x}_{k-1}^+ + G_{k-1} \hat{u}_{k-1} \label{eqn:covA} \\
{P}_k^- & = &F_{k-1} {P}_{k-1}^+ F_{k-1}^T + Q_{k-1}\label{eqn:covB}\\
q_k & = &y_k - H_k \hat{x}_k^- \label{eqn:covC}\\
{S}_k & = &H_{k} {P}_{k}^- H_{k}^T + R_{k}\label{eqn:covD}\\
{K}_k & = &{P}_{k}^- H_{k}^T(H_{k} {P}_{k}^- H_{k}^T + R_{k})^{-1}= {P}_{k}^- H_{k}^T {S}_{k}^{-1}\label{eqn:covE}\\
\hat{x}_k^+ & = &\hat{x}_k^- + K_{k} q_k \label{eqn:covF} \\
{P}_k^+ & = &(I - K_k H_k) {{P}_k^-} (I - K_k H_k)^T + K_k R_k K_k^T \nonumber \\
& = &(I - K_k H_k) [F_{k-1}{P}_{k-1}^+ F_{k-1}^T + Q_{k-1}] \nonumber \\
& & (I - K_k H_k)^T + K_k R_k K_k^T \label{eqn:covG}
     \end{eqnarray}

For the design of a suitable KF, four tuning parameters need to be
determined prior to implementing the filter. These are the initial
state estimate $x_{0}$ and the three uncertainty or noise
covariance matrices, namely the initial state (estimation) error
covariance $P_0$, the process (model) noise covariance $Q$ and the
measurement noise covariance $R$. Of these covariances, $P_0$ is
only the initial choice of the state estimation error covariance $P_k$, which has to be decided by the designers, since the state estimation
error covariance $P_k$ changes as the filtering progresses with time and is
expected to reach a steady value as the filter converges, provided
that the system is not subjected to any major change in the system
input. However, $Q$ and $R$ have to be decided for the total duration of filtering and depending on practical considerations, designers choose these to be
time-invariant or time-varying.

The choice of the elements of $x_0$ and $P_0$ may range from
infinitely large values to small values depending on the available
information about the relevant states \cite{Zarchan2000}. It has
further been observed that the choices of $x_0$ and $P_0$ mainly
affect the initial part of the estimation so these are usually not
very critical unless the initial estimation exceeds acceptable
limits \cite{Zarchan2000}. The choice of $R$ is also non-critical
since the sensor characteristics, which are usually known
beforehand, can be used to decide on a suitable matrix.

 Among all the tuning parameters, the tuning of the process noise covariance $Q$
 is considered to be the most critical \cite {Ananthasayanam2011}. This is so since
  all the model uncertainties and inaccuracies as well as the noises affecting the process
  are incorporated quantitatively into $Q$. It is also known that a \emph{proper} ratio of the
  filter tuning parameter values affects the filter performance \cite {Shalom2001,Zarchan2000}.
  This further complicates the choice of a suitable $Q$.

In the present work, the innovation $q_k$, which directly affects
the Kalman gain, has been identified as the critical parameter
which can be utilized to predict the proper choice of $Q$ for the
filter, given an arbitrary choice of $x_0$, $P_0$ and $R$, which
may or may not be identical to the values obtained from the actual
system. This use of $q_k$ can be justified from the established
fact that the Kalman gain plays a major role in ensuring the
optimized performance of the KF  while the NIS ensures filter
consistency \cite {Ananthasayanam2011, Ananthasayanam2002,
Shalom2001}. However, the innovations, which are random variables obtained in real-time, are
not quite helpful for predictions due to lack of a deterministic
basis.

So, in order to obtain a deterministic basis for the prediction of the \emph{suitable} filter tuning parameters, the innovation error covariance $S_k$, as stated in eqn ({\ref{eqn:covD}), is studied instead. It is observed that $S_k$ depends on the apriori estimation error covariance \emph{for the measured outputs} (not states), namely $H_k P_k^- H_k^T$.

Let the innovation covariance $S_k$ in eqn ({\ref{eqn:covD}) be expressed as
\begin{equation}\label{eqn:Sk}
S_k = H_k (F_{k-1}{P}_{k-1}^+ F_{k-1}^T +Q_{k-1})H_k^T+R_k=(A_k+B_k+R_k)
\end{equation}
where $A_k=H_k F_{k-1}{P}_{k-1}^+ F_{k-1}^T H_k^T$ and $B_k=H_k Q_{k-1} H_k^T$.

So, using the expressions for the Kalman gain $K_k$, the apriori state estimation error covariance $P_k^-$ and the innovation covariance $S_k$ from eqns ({\ref{eqn:covE}), ({\ref{eqn:covB}) and ({\ref{eqn:Sk}) respectively, we obtain
\begin{equation}\label{eqn:HK}
H_k K_k = H_k P_k^- H_k^T S_k^{-1} = (A_k+B_k) (A_k+B_k+R_k)^{-1}.
\end{equation}

Thereafter, the aposteriori state estimation error covariance $P_k^+$, as stated in eqn ({\ref{eqn:covG}), is pre- and post-multiplied by $H_k$ and $H_k^T$ respectively, to obtain
\begin{multline}\label{eqn:HkPkplusHkT}
    H_k {P_k^+} H_k^T =[H_k (F_{k-1} {P}_{k-1}^+ F_{k-1}^T)H_k^T ]\\
   + B_k-(A_k+B_k)(A_k+B_k+R_k)^{-1}(A_k+B_k).
\end{multline}
 Pre-multiplying both sides of the previous equation by $(A_k+B_k)^{-1}$ and rearranging the terms yields
\begin{multline}\label{eqn:preJ1J2}
   (A_k+B_k)^{-1} [H_k (P_k^+ - F_{k-1} {P}_{k-1}^+ F_{k-1}^T)H_k^T ]\\
   = (A_k+B_k)^{-1}B_k-(A_k+B_k+R_k)^{-1}(A_k+B_k)\\
      = (A_k+B_k)^{-1}B_k+(A_k+B_k+R_k)^{-1}R_k - I_m.
\end{multline}
 Taking the trace of both sides of eqn (\ref{eqn:preJ1J2}) and rearranging, we obtain the two new metrics $J_{1k}$ and $J_{2k}$ as
\begin{eqnarray}\label{eqn:J1kJ2k}
   J_{1k} + J_{2k}& = &m- trace\{N_k\}
\end{eqnarray}
where
\begin{eqnarray*}
{\begin{array}{lll}
J_{1k}& =& trace\{(A_k+B_k+R_k)^{-1}R_k \}\\
 J_{2k}& = &trace\{(A_k+B_k)^{-1}B_k \}\\
  $and$ \; N_k &=&(A_k+B_k)^{-1} [H_k (F_{k-1} {P}_{k-1}^+ F_{k-1}^T-P_k^+)H_k^T].
  \end{array}}
  \end{eqnarray*}
 It is useful to note from eqn (\ref{eqn:J1kJ2k}) that the value of $J_{1k}+J_{2k}$ at any instant $k$ deviates
 from the number of measurements $m$ due to the contribution of the term $trace\{N_k\}$ at that particular instant of time $k$.

For the evaluation of the overall filter performance, let the performance indices or metrics $J_1$ and $J_2$ and a controlling parameter for the metrics, $n_q$, be defined in terms of the total horizon $N$ as
\begin{eqnarray} \label{eqn:J1J2}
    J_{1} & = & \frac{1}{N}\sum_{k=1}^N J_{1k} \nonumber \\
    & = & \frac{1}{N}\sum_{k=1}^N trace\{(A_k+B_k+R_k)^{-1}R_k \} \nonumber \\
    J_{2} & = & \frac{1}{N}\sum_{k=1}^N J_{2k} \nonumber \\
    & = & \frac{1}{N}\sum_{k=1}^N trace\{(A_k+B_k)^{-1}B_k \} \nonumber\\
    \mbox{and} \; n_{q} & = & \frac{1}{N}\sum_{k=1}^N log\{ trace(B_k
    )\}.
\end{eqnarray}

In order to appreciate the significance of the two new metrics, it must be noted that the focus of the present work is on the
\emph{predicted measurement} and the factors contributing to it while the standard treatment in the existing literature
\cite{Shalom2001,Brown1996,Simon2006} focuses simply on the estimated states and the errors thereof.

 It is to be noted that the apriori state estimation covariance $F_k P_{k-1}^+ F_k^T$ and the process noise covariance $Q_{k-1}$ have been projected from the state equation onto the innovation covariance $S_k$ as $A_k$ and $B_k$ respectively, as evident from eqn (\ref{eqn:Sk}). So, $J_{2k}$ provides a measure of the effect of the independent parameter, namely the process noise covariance $Q_{k-1}$, and specifically, its projection onto the measurement, namely $B_k$, on the sum $(A_k+B_k)$. This sum forms the state dependent component of the innovation covariance $S_k$. $J_{1k}$, on the other hand, provides the effect of the other independent parameter, namely the measurement noise covariance $R_k$, which is directly related to the measurement, on the total sum $(A_k+B_k+R_k)= S_k$. Since the innovation covariance affects the value of the Kalman gain and hence the correction in the state estimate, so the metrics $J_{1k}$ and $J_{2k}$ provide a predictive insight into the proper choice of the tuning parameters, as seen from the observations listed hereafter.

 To evaluate the performances of the KF and the EKF in a particular application, the four filter tuning parameters $x_0, P_0^+, R$
may be fixed apriori, as stated earlier in this Section, while $Q_0$ is continuously varied. A change in $Q_0$
changes the overall $n_q$, irrespective of the choice of a fixed or variable $Q_k$.
Furthermore, this change in $Q_0$ affects the general tuning parameter ratios $P/Q$ and $Q/R$ at all instants and
for all the states. As stated in \cite{Shalom2001}, these affect the overall filter performances.

 The following are observed from eqn (\ref{eqn:J1J2}) for a changing $Q_0$, and hence a changing $n_q$.

\begin{itemize}
\item[(i)]For very large $Q_{0}$ in terms of the trace, $B_k$ is much larger than $R$ in terms of the trace, and so, $J_1 \; \rightarrow \; 0$. \\
Also in this case, $J_2$ tends to the number of measurements $m$ for a convergent and small $A_k$ and typically reaches a steady value.
\item[(ii)]Similarly, when $Q_0$ is small in terms of the trace, then $B_k$ is significantly smaller than $R$ in terms of the trace, so $J_1 \; \rightarrow \; m$ and typically reaches a steady value while $J_2 \; \rightarrow \; 0$.
\item[(iii)]When $B_k$ is comparable to $R$ in terms of the trace, then $J_1$ and $J_2$ change significantly from both the upper and lower bounds stated earlier.
\end{itemize}

It is thus observed that when $n_q$ changes from a small value to a large value, $J_1$ decreases from $m$ to $0$, while $J_2$ increases from $0$ to $m$. So, the natures of the change of the metrics $J_1$ and $J_2$ are contrary to each other and hence, it is to be expected that there will be a crossover of the two plots $J_1$ vs. $n_q$ and $J_2$ vs. $n_q$ for some value of $n_q$ and hence some value of $Q_0$, henceforth denoted as $Q_{comp}$. The change in $J_1$ can be considered to be driven by $R_k$ while the change in $J_2$ is driven by $B_k$. So, a higher value of the metric $J_1$ improves the sensitivity of the filter performance while an increase in the metric $J_2$ improves the filter robustness. Thus, a compromise, or a balance, between these two factors, namely sensitivity and robustness, can be predicted in the RMSE performance by choosing a suitable value of $Q_0$ close to $Q_{comp}$.

\section{Case Study}
 In order to verify the predictability of the performance of a filter for different combinations of the tuning parameters using the
 defined metrics $J_1$ and $J_2$, the problem considered in the present work is that of the tracking of a 2D ballistic target as
 discussed in \cite{Farina2002}. It is assumed that the object enters the atmosphere in reentry phase under the presence of
 nonlinear air drag as well as gravity and so, the trajectory of the target is a nonlinear one.

The equivalent discrete-time target motion can be expressed using the nonlinear state equation \cite{Farina2002}
\begin{eqnarray} \label{eqn:case-state}
x_{k+1}& = & f(x_k) + Gu_k +w_k \nonumber \\
        & =& F x_k + G f_{kk}(x_k) + Gu_k +  w_k
\end{eqnarray}
where the state vector
$x_k = [x_{1k} \: \dot{x}_{1k} \: x_{2k} \: \dot{x}_{2k}]^T $ consists of the positions in the $x$ and $y$ directions denoted as
$x_{1k}$ and $x_{2k}$ respectively and their corresponding velocities.  $u_k=[0 \; (-g)]^T$ and $w_k=N(0,\sqrt{Q_t})$ are the input
matrix and the process noise respectively.

The function $f_{kk}(x_k) = \; -\frac{1}{2\beta}\rho g \sqrt{{\dot{x}_{1k}}^2
+ {\dot{x}_{2k}}^2}\; [\dot{x}_{1k} \: \dot{x}_{2k}]^T$, while

\begin{equation*}
Q_t=
\left[{\begin{array}{cccc}
 4 & 2 & 0& 0\\
  2 & 2 &0  &0\\
   0 & 0 & 4& 2\\
    0& 0 & 2&2
\end{array} }\right].
\label{matrix_RA}
\end{equation*}
These and all other terms, as used in the present problem, are as defined
in \cite{Farina2002}. Only, the target ballistic co-efficient $\beta$ has been assumed to be known with a constant value of $\beta=40000 kgm^{-1}s^{-2}$. The acceleration due to gravity $g$ is assumed to be constant at $9.81ms^{-2}$.

The ballistic target trajectory has been generated considering the initial value of the state vector as $x_0 = [232km  \; \; 2.29cos(190^{\circ})kms^{-1} \; \; 88km \; \;  2.29sin(190^{\circ})kms^{-1}]^T$
and the air density function $\rho = C_1e^{-C_2x_2}$ with \\
$C_1=1.227 , \; C_2=1.093\times10^{-4}$ when $x_2<9.144km$ \\
and $C_1=1.754 , \; C_2=1.490\times10^{-4}$ when $x_2>9.144km$.\\
This change in the air density during the reentry becomes very crucial during the estimation as the system nonlinearity changes abruptly at this height.

The measurement equation is considered to be
\begin{equation}\label{eqn:case-meas}
y_{mk}= h(x_k)+v_k
\end{equation}
where the terms have their usual meaning and the measurement noise is considered to be $v_k=N(0,\sqrt{R_A})$. \\
In the practical scenario, the measurements available depend on the choice of the sensor(s) and other practical constraints.
In order to obtain the measurements of the positions $x_{1m}$ and $x_{2m}$, the simulated radar measurements of range  $r$
and angle $\varepsilon$, available in polar co-ordinates, are converted into Cartesian coordinates. The relations used
are $x_{1m} = rcos(\varepsilon)$ and $x_{2m} = rsin(\varepsilon)$. These data are further corrupted with the randomly
generated zero-mean measurement noise $v_k$ using the noise covariance matrix $R_A$  given as

\begin{equation}
R_A=
\left[{\begin{array}{cc}
\sigma^2_d & \sigma_{dh}\\
\sigma_{dh} & \sigma^2_h
\end{array} }\right]
\label{matrix_RA}
\end{equation}
where $\sigma_r =100m$ is the variance of $r$, $\sigma_{\varepsilon} ={(.017/57.3)}^{\circ}$ is the variance of $\varepsilon$ and
\begin{eqnarray}
\sigma_d^2 & = & \sigma_r^2 \: cos^2(\varepsilon)+ r^2 \: \sigma_{\varepsilon}^2 \: sin^2(\varepsilon) \nonumber \\
\sigma_h^2 & = & \sigma_r^2 \: sin^2(\varepsilon)+ r^2 \: \sigma_{\varepsilon}^2 \: cos^2(\varepsilon) \nonumber \\
\mbox{and} \; \sigma_{dh} & = & (\sigma_r^2 - r^2 \: \sigma_{\varepsilon}^2 ) \: sin(\varepsilon)\; cos(\varepsilon). \nonumber
\end{eqnarray}

\section{Simulation and Results}
 In order to evaluate the performances of the KF and the EKF, the four filter tuning parameters $x_0, P_0^+, R$ and a nominal $Q_0$,
 henceforth referred to as $Q_{nom}$,  have to be fixed. In this case, $x_0$ and $P_0^+$ are obtained using the two point
 differencing method as stated in \cite{Shalom2001} and \cite{Farina2002},  $R$ is  obtained from the sensor and is considered
 to be time-invariant while $Q_{nom}$ is obtained as the adaptive $Q$ as discussed in \cite{Ananthasayanam2002}.
 The values of these tuning parameters, which are considered to be the same for both KF and EKF, are as follows:
\begin{eqnarray*}
 x_0 \; = &
 \left[{\begin{array}{c}
2.25 \times 10^5   \\
-2.81 \times 10^3 \\
 9.26 \times 10^4 \\
6.75 \times 10^3
\end{array} }\right], 
R \; = &
 \left[{\begin{array}{cc}
10.54 & -3.85   \\
-3.85 & 37.15
\end{array} }\right] \\
\end{eqnarray*}
\begin{multline*}
P_0^+  = \\
\left[{\begin{array}{cccc}
2.48 \times 10^6  & 0    & -6.76 \times 10^6    & 0 \\
0     & 1.24 \times 10^6  & 0    & -1.73 \times 10^6 \\
-6.76 \times 10^6   & 0    & 1.47 \times 10^7 & 0 \\
0     & -0.73 \times 10^6    & 0    & 7.34 \times 10^6
\end{array} }\right]\\
\end{multline*}
\begin{multline*}
Q_{nom} = \\
\left[{\begin{array}{cccc}
2.48 \times 10^5  & 6.32 \times 10^4   & -5.10 \times 10^5    & -1.04 \times 10^5 \\
6.32 \times 10^4     & 2.34 \times 10^4  & -1.04 \times 10^5   & -2.88 \times 10^4 \\
-5.10 \times 10^5   & -1.04 \times 10^5    & 1.44 \times 10^6 & 3.45 \times 10^5 \\
-1.04 \times 10^5     & -2.88 \times 10^4    & 3.45 \times 10^5    & 1.20 \times 10^5
\end{array} }\right]
 \end{multline*}
In order to predict the performances of the KF and the EKF  and compare them for different combinations of the tuning parameters,  the metrics $J_1$ and $J_2$ have been obtained for varying $n_q$ as shown in Fig.\ref{fig_J1_J2_A} and Table \ref{table_caseA} by using $Q_0= 10 ^{p} Q_{nom}$, and varying $p$ suitably, as evident from the first two columns of the Table. These predictive metrics have been compared with the RMSE performances of the KF, (Fig.\ref{fig_RMSE_KF_A}), and the EKF, (Fig.\ref{fig_RMSE_EKF_A}), obtained using the same tuning parameters.

\begin{figure}[h]
\begin{center}
\includegraphics[width=3.5in]{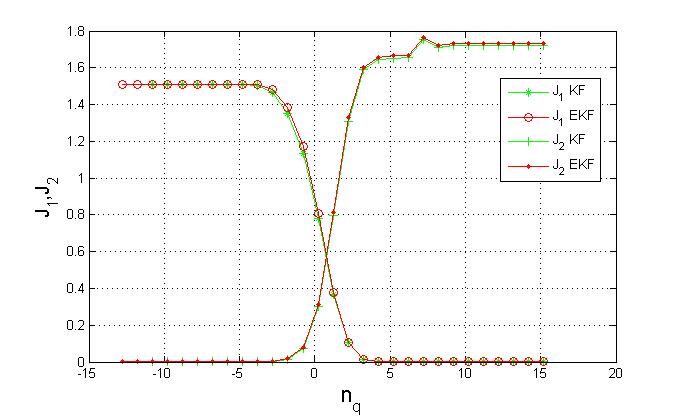}
\caption{Plots of $J_1$ and $J_2$ vs. $n_q$ for KF and EKF}
\label{fig_J1_J2_A}
\end{center}
\end{figure}

\begin{table}[h]
\centering
\caption{Metrics $J_1$ and $J_2$ for KF and EKF}
\label{table_caseA}
\begin{tabular}{|c|c|c|c|c|c|c|c|c|c|c|}
\hline  $p$ & $n_q$ & $J_1$ KF & $J_2$ KF & $J_1$ EKF & $J_2$ EKF \\
\hline -13 & -6.79 & 1.51  & 0.00  & 1.51 & 0.00 \\
\hline -12 & -5.79 & 1.51  &  0.00 &  1.51 &  0.00  \\
\hline -11 & -4.79 & 1.51  &  0.00 &  1.51  &  0.00  \\
\hline -10 & -3.79 & 1.50  &  0.00 &  1.51  &  0.00  \\
\hline -9 & -2.79 & 1.46  &  0.00 &  1.48  &  0.00  \\
\hline -8 & -1.79 & 1.35  &  0.01 &  1.39  &  0.02  \\
\hline -7 & -0.79 & 1.13  &  0.07 &  1.17  &  0.08  \\
\hline  -6 & 0.21 & 0.78  &  0.30 &  0.81  &  0.31  \\
\hline  -5 & 1.21 & 0.37  &  0.80 &  0.38  &  0.81  \\
\hline  -4 & 2.21 & 0.10  &  1.31 &  0.11  &  1.33  \\
\hline  -3 & 3.21 & 0.02  &  1.59 &  0.02  &  1.60  \\
\hline  -1 & 4.21 & 0.00  &  1.65 &  0.00  &  1.66  \\
\hline  0 & 5.21 & 0.00  &  1.65 &  0.00  &  1.66  \\
\hline  1 & 6.21 & 0.00  &  1.65 &  0.00  &  1.67  \\
\hline  2 & 7.21 & 0.00  &  1.75 &  0.00  &  1.76  \\
\hline  3 & 8.21 & 0.00  &  1.71 &  0.00  &  1.72 \\
\hline  4 & 9.21 & 0.00  &  1.72 &  0.00  &  1.73 \\
\hline  5 & 10.21 & 0.00 &  1.72 &  0.00  &  1.73  \\
\hline  6 & 11.21 &  0.00 & 1.72 &  0.00  &  1.73 \\
\hline
\end{tabular}
\end{table}

 As can be predicted from the derivations, it is observed from Fig.\ref{fig_J1_J2_A} and Table \ref{table_caseA} that, for changing $Q_0$ and hence, for changing $n_q$,

\begin{itemize}
\item[1)]Both $J_1$ and $J_2$ are bounded by the number of measurements $m=2$ in the upper limit and $0$ in the lower limit for significantly high and low values of $n_q$.  As $n_q$ changes, the values of both $J_1$ and $J_2$ change  between these limits of $m$ and 0. 
\item[2)]While $P_0$ and $R$ are fixed, $J_1$ attains a steady maximum value close to $m$ for low $n_q$ while $J_2$ also reaches a steady maximum value close to $m$, but for high $n_q$. In this case, it is observed that $J_1$ attains a steady value of $1.51$, for KF as well as EKF, for values of $n_q$ less than $-4.79$ and $-3.79$ respectively while $J_2$ attains a steady value of $1.72$ and $1.73$ for KF and EKF respectively, for $n_q$ larger than $9.21$ in both the cases. A minimum value, very close to 0, is obtained for $J_2$ when $n_q$ is very small, typically -2.79 or less, and for $J_1$, when $n_q$ is very large, typically 4.21 or more, for both KF and EKF. 
\item[3)] At a particular value of $n_q$, when the corresponding value of $Q_0$ is the compromise value, $Q_{comp}$, there is a crossover of the plots for $J_1$ and $J_2$. In this case, the corresponding value of $n_q$ for $Q_{comp}$ lies between $0.21$ and $1.21$ for both KF and EKF. 
\item[4)]The filter exhibits robustness for those combinations of the tuning parameters for which the value of $J_2$ is close to the number of measurements $m$. On the other hand, when the value of $J_1$ is close to the number of measurements, this indicates a sensitivity in the RMSE performance but this might also cause the filter to diverge if the actual system noise or disturbances are large. Hence, it can be said that the robustness of the RMSE  performances increase for KF and EKF for $n_q \geq 1.21$ while sensitivity of the RMSE performances of the filters increase for $n_q \leq 0.21$. This is validated from the filter performances as observed in Figs. \ref{fig_RMSE_KF_A} and \ref{fig_RMSE_EKF_A}. 
\item[5)]Identical values of the metrics obtained using the same filter indicate similarity of the RMSE performances. From the RMSE plots for KF (Fig.\ref{fig_RMSE_KF_A}) and EKF (Fig.\ref{fig_RMSE_EKF_A}), it is observed that similar robustness in the RMSE performances are obtained for $n_q >4.21$ while similar sensitive nature of the RMSE  performances are obtained for $n_q < (-4.79)$. For the KF filter, the position estimates show lower RMSE but with a tendency of divergence during the end phase whereas the velocity estimates have additional initial spikes in the sensitive zone of $n_q$ while the RMSE performances in this zone are much improved in the case of the EKF. However, it is interesting to note that the performances of both the KF and the EKF filters are quite similar in the robust zone. It is also observed that at $n_q=7.21$, the metric $J_2$ for both the filters shows a positive spike, with values of $1.75$ and $1.76$ for KF and EKF respectively. From the RMSE plots, it is observed that this corresponds to high initial peaks in the position estimates in both filters.
\item[6)]In the zone where the values of $J_1$ and $J_2$ are changing and specifically near the crossover point, the trade-off between sensitiveness and robustness of the filter can be achieved with a judicious choice of the combination of the tuning parameters. A non-judicious choice may lead to filter instability. So, the best compromise in the RMSE performances is expected near $Q_0=Q_{comp}$. This is validated from the RMSE plots for both filters for $0.21 \leq n_q \leq 1.21$, where there are no sharp peaks in the initial phase nor is there any divergent behaviour at the end phase.
\end{itemize}

\begin{figure}[h]
\begin{center}
\subfigure[]{\includegraphics*[width=2.5in,height=2.0in]{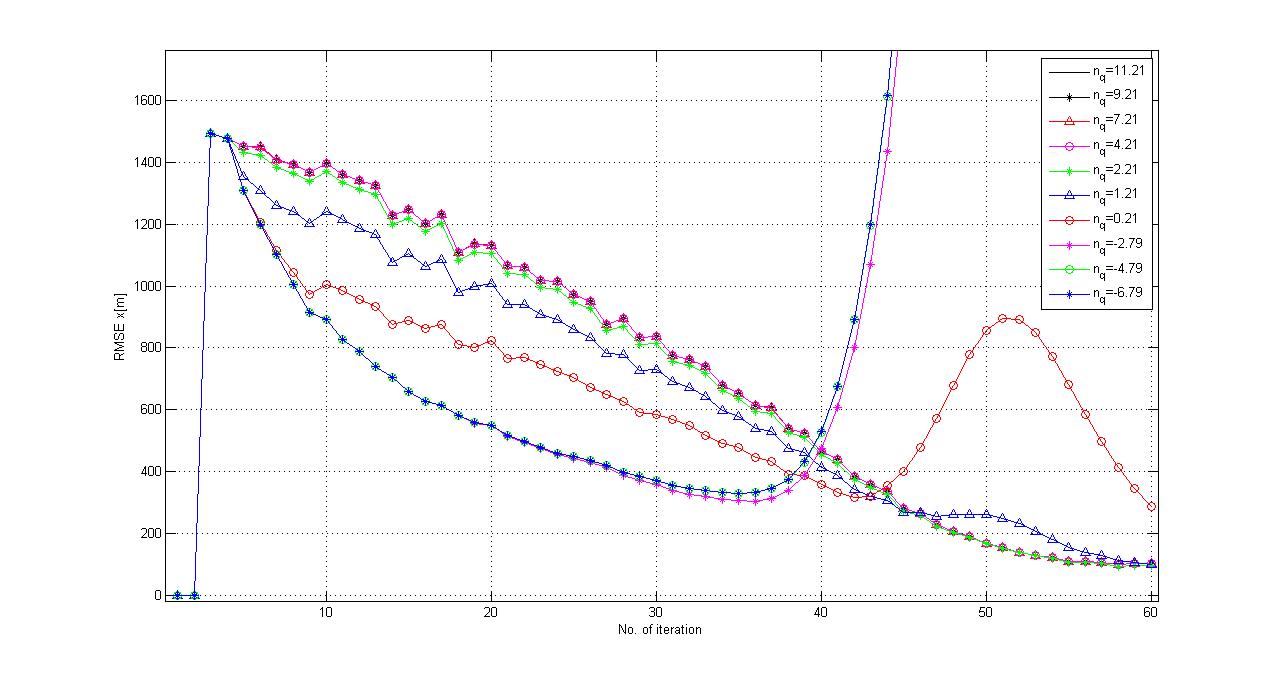}}
\subfigure[]{\includegraphics*[width=2.5in,height=2.0in]{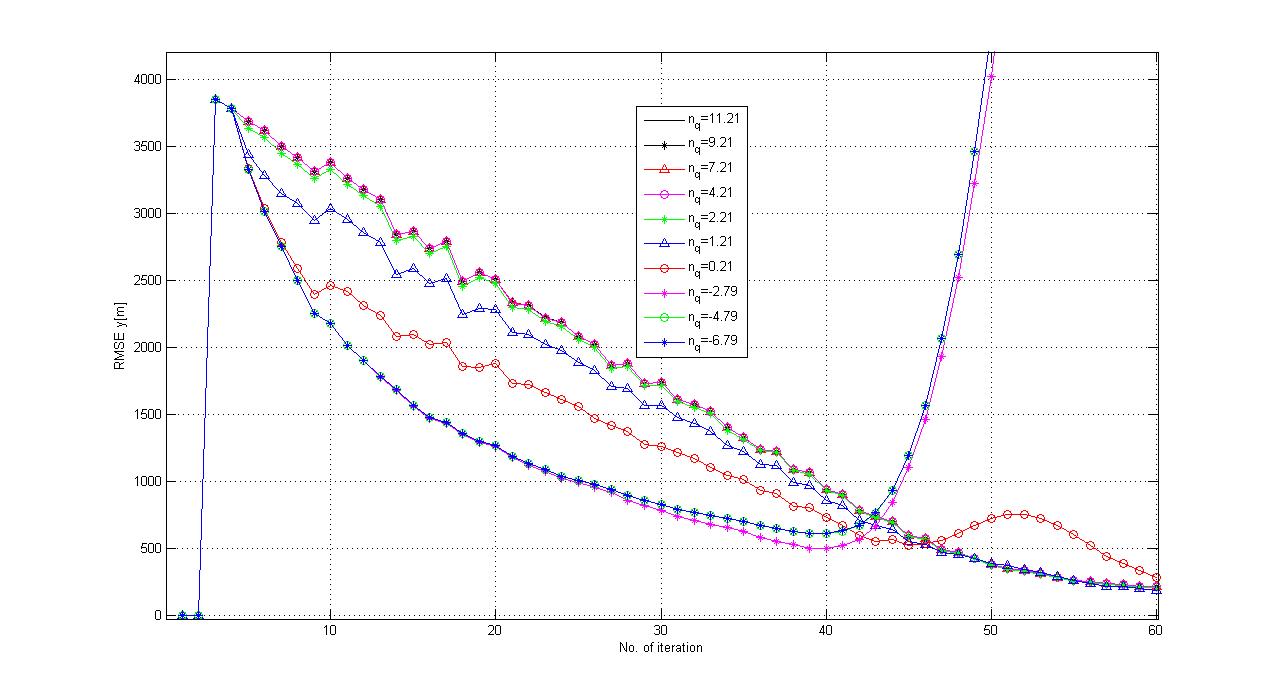}}\\
\subfigure[]{\includegraphics*[width=2.5in,height=2.0in]{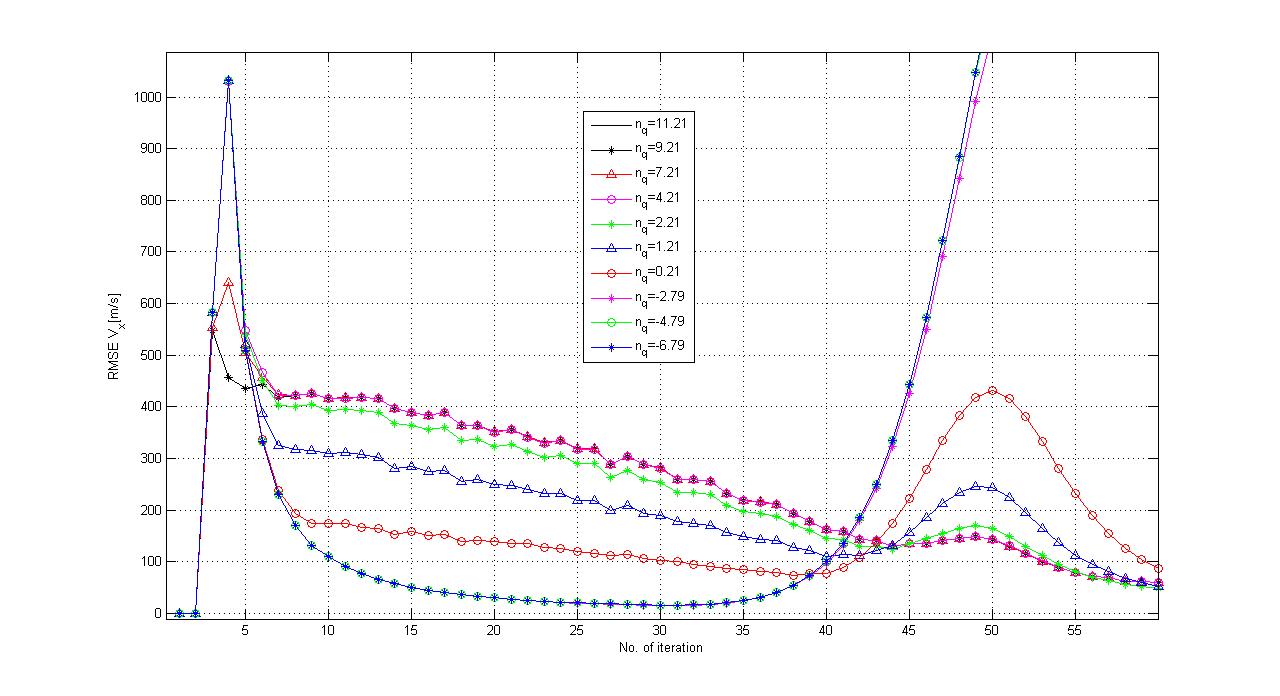}}
\subfigure[]{\includegraphics*[width=2.5in,height=2.0in]{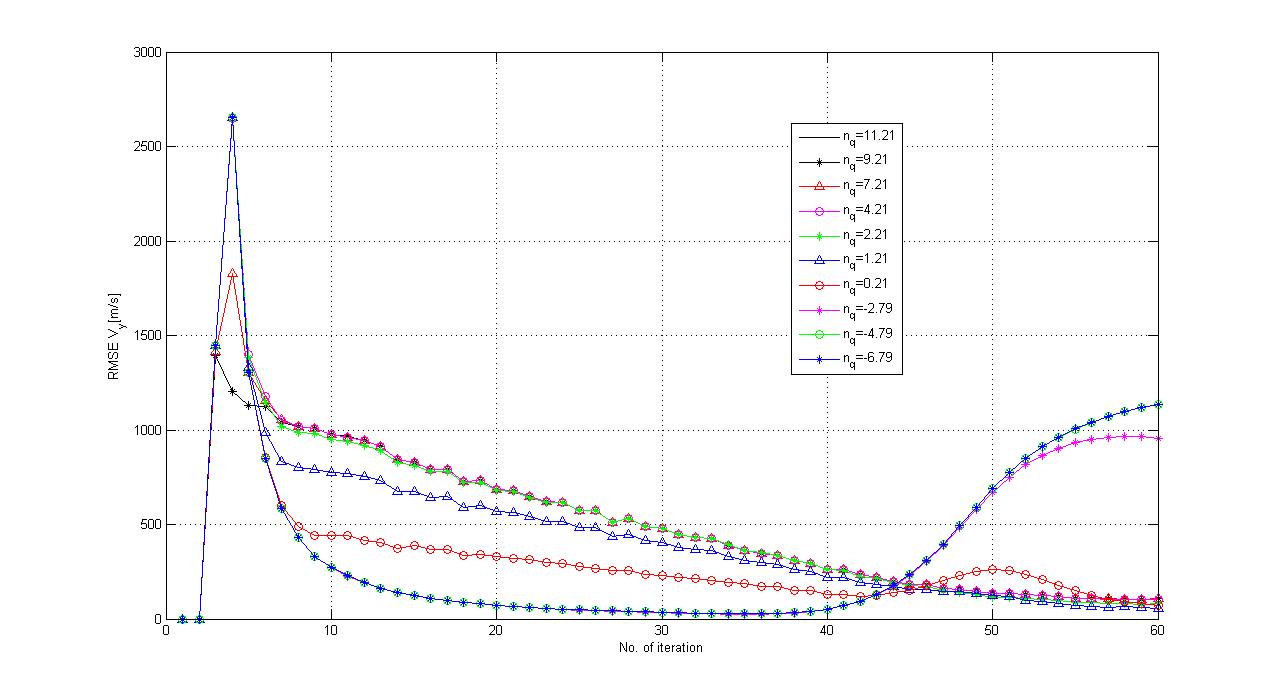}}
\caption{RMSE plots of the positions a)$x$ and b)$y$ and velocities c)$V_x$ and  d)$V_y$ using KF}
\label{fig_RMSE_KF_A}
\end{center}
\end{figure}

\begin{figure}[h]
\begin{center}
\subfigure[]{\includegraphics*[width=2.5in,height=2.0in]{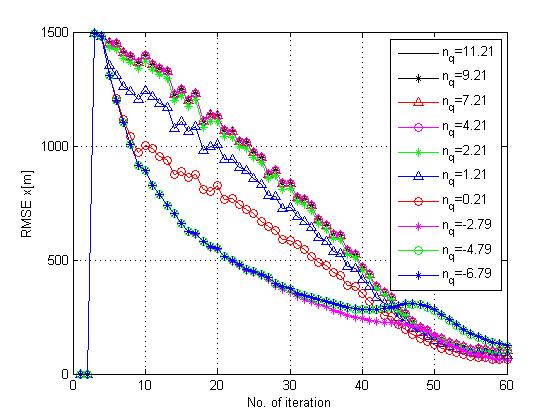}}
\subfigure[]{\includegraphics*[width=2.5in,height=2.0in]{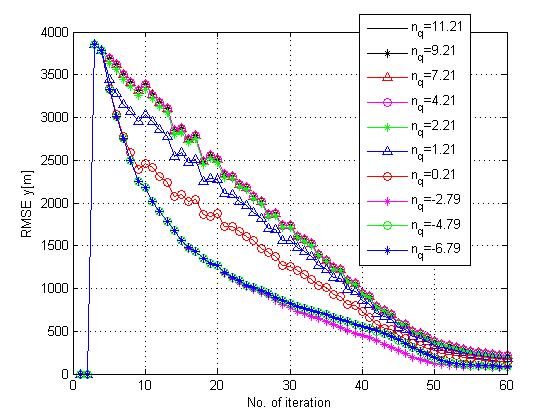}}\\
\subfigure[]{\includegraphics*[width=2.5in,height=2.0in]{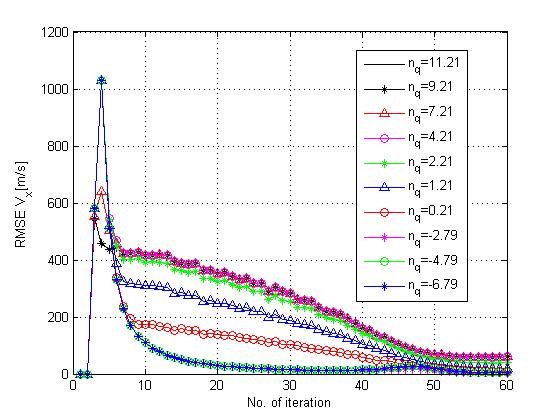}}
\subfigure[]{\includegraphics*[width=2.5in,height=2.0in]{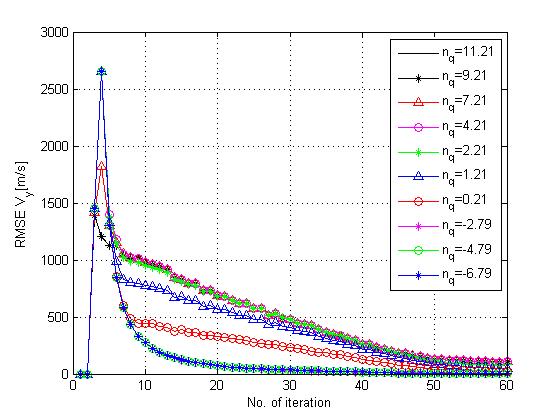}}
\caption{RMSE plots of the positions a)$x$ and b)$y$ and velocities c)$V_x$ and  d)$V_y$ using EKF}
\label{fig_RMSE_EKF_A}
\end{center}
\end{figure}

\section{Conclusion}
In the present work, two new metrics $J_1$ and $J_2$ have been proposed for determining the filter tuning parameters on the basis of the innovation covariance $S_k$ by using the concept of the  \emph{predicted measurement} and the factors contributing to it. It is to be noted that the standard treatment in the existing literature \cite{Shalom2001,Brown1996,Simon2006} focuses simply on the estimated states and the errors thereof. So, the proposed approach provides a major shift in the filtering paradigm.

For predicting the proper combination of the filter tuning parameters, these performance indices have been evaluated for a 2D ballistic target problem \cite{ Farina2002} as shown in Fig.\ref{fig_J1_J2_A} and Table \ref{table_caseA}. For this, the critical tuning parameter is the process noise covariance matrix $Q$ \cite{Ananthasayanam2011}. This matrix has been tuned in a continuous manner by considering $Q_0=10^{p} Q_{nom}$ where $Q_{nom}$ is any suitable  nominal choice of the $Q$ matrix and the multiplier $p$ is varied continuously. It is to be noted that this,in effect, also varies a controlling parameter $n_q$, as defined in eqn. (\ref{eqn:J1J2}). All the other filter tuning parameters, $x_0$, $P_0^+$ and $R$, are kept fixed. The actual filter performances can be predicted from the nature of change of the metrics $J_1$ and $J_2$ with the change of $n_q$, which can both be calculated offline. These predictions have been validated in terms of the RMSE performances for both the KF (Fig.\ref{fig_RMSE_KF_A}) and EKF (Fig.\ref{fig_RMSE_EKF_A}) and are summarized hereafter.

It is observed that both $J_1$ and $J_2$ are bounded by the number of measurements $m=2$ in the upper limit and $0$ in the lower limit for significantly high and low values of $n_q$.  As $n_q$ changes, the values of both $J_1$ and $J_2$ change  between these limits of $m$ and 0.
While $P_0$ and $R$ are fixed, $J_1$ attains a steady maximum value close to $m$ for low $n_q$ while $J_2$ reaches a maximum value close to $m$ for high $n_q$.
A minimum value, very close to 0, is obtained for $J_2$ when $n_q$ is very small and for $J_1$, when $n_q$ is very large, for both KF and EKF.

At a particular value of $n_q$, when the corresponding value of $Q_0$ is the compromise value, $Q_{comp}$, there is a crossover of the plots for $J_1$ and $J_2$. In this application, the corresponding value of $n_q$ for $Q_{comp}$ lies between $0.21$ and $1.21$ for both KF and EKF.
A value of $J_2$ close to the number of measurements $m$ indicates that the filter exhibits robustness for those combinations of the tuning parameters. On the other hand, when the value of $J_1$ is close to the number of measurements, this indicates a sensitivity in the RMSE performance but this might also cause the filter to diverge. Hence, robustness of the RMSE performances are expected and obtained for KF and EKF for $n_q \geq 1.21$ while sensitivity of performances of the filters increase for $n_q \leq 0.21$. However, there is a limit to both of these since identical metrics for the same filter yields identical RMSE performances. So, the proposed metrics $J_1$ and $J_2$ can be used by the design engineer to decide suitable choices of the filter tuning parameters based on the performance requirements for the system.

Further studies have been and are being performed by the present researchers using these predictive metrics for different linear and nonlinear system and measurement scenarios which validate the observations stated in this work. These metrics and their comparison with the corresponding NIS values, are expected to provide additional insight into the choice of the filter tuning parameters and their effects on the RMSE performances.

\section*{Acknowledgements}
The authors would like to thank Prof. M. K. Ghosh, Department of Mathematics, IISc, Bangalore for his active help, critical comments and insights into the derivations. They would also like to thank Dr. A. Sarkar, DRDO, Hyderabad, Prof. T. K. Ghoshal, Jadavpur University, Kolkata and Prof. M. Ananthasayanam, IISc, Bangalore for their help in emphasising the various aspects of Kalman filtering. The authors thank the anonymous reviewers for their insightful comments and suggestions.

\bibliographystyle{IEEEtran}
\bibliography{mybibfile_bg}

\end{document}